\documentclass[10pt,a4paper,twoside]{article}

\usepackage{indentfirst}            
\usepackage{graphicx}               
\usepackage{amsgen,amsfonts,amssymb,amsbsy} 
\usepackage{amsmath}%

\newenvironment{resum}{\begin{quote}\small}{\end{quote}}

\newcommand{\bfsf}[1]{\textsf{\textbf{#1}}}

\begin{document}

\thispagestyle{plain}       

\begin{center}


{\LARGE\bfsf{LABORATORY PHYSICS AND\\[0.25cm] COSMOLOGY\footnote{%
Accepted for publication in \emph{Physics Essays}\/ in 2004.}  }}

\bigskip


\textbf{Antonio Alfonso-Faus }


\emph{E.U.I.T. Aeron\'autica,  Spain\\ Plaza Cardenal Cisneros s/n \\ MADRID 28040, SPAIN }\/

\end{center}

\medskip


\begin{resum}
Abstract

Following four different fundamental principles, in addition to  Weinberg's universal
relation, we find five reasons demanding that any gravitational mass $m$, and the speed of light
$c$, vary with cosmological time (the masses increasing linearly with time and the speed of
light decreasing with time, in such a way that the product $mc$ remains constant). This is
required by the universal condition of conservation of momentum in a Universe with spatial
homogeneity: a consequence of the Milne-Einstein COSMOLOGICAL PRINCIPLE. We prove that this is
consistent with Einstein's Theory of General Relativity (through the application of the Action
Principle). We call this effect a ``MASS BOOM''. At the LAB system no such time variations can
be detected, unless we make comparisons with cosmological observations (i.e. using different
references far away from the LAB, e.g. measuring the red shift from distant galaxies in terms
of the LAB standards). We have to stress that the physical conditions implied by a time
varying mass, like the Mass Boom establishes, together with a time varying speed of light,
preserving the constancy of momentum, are compatible with Einstein's field equations. We then
integrate his cosmological equations and find the solution for the cosmological scale factor
as $a(t) = \text{constant} \cdot t^2$, implying an apparent accelerated expansion for the
Universe, as seen from the LAB frame. This is the interpretation given to recent observations
obtained from the Supernova Type Ia. The determination of the scale factor as $a(t) =
\text{constant} \cdot t^2$ is based upon a LAB interpretation and therefore is an apparent
effect. On the other hand we note that the product $ct$ being a constant determines the real
Universe as a static one, of constant size, in accordance with Einstein´s first proposal. The
observed red shift at the LAB system is due to a real shrinkage of the quantum world, due to
the decrease in size of the quantum particles determined by a decreasing Planck's ``constant''
$\hbar$ (quantum sizes are of order $\frac{\hbar}{mc}$, where $mc$ remains constant).

\bigskip\noindent
Key words: Cosmology, Action Principle, fundamental physical constants, gravity quanta,
supernova, Mach's principle, General Relativity.

\bigskip

\end{resum}

\begin{resum}
R\'esum\'e

On se base en quatre principes fondamentaux diff\'erents, et en plus la relation universelle
de Weinberg, on trouve cinq raisons qui exigent que n'importe quelle masse de la gravit\'e
$m$, et la vitesse de la lumi\`{e}re $c$, changent avec le temps cosmologique (les masses
augmentant lin\'eairement avec le temps et la vitesse de la lumi\`{e}re diminuant avec le
temps, de telle mani\`{e}re que le produit $mc$ reste constant). Ceci est exig\'e par la
condition universelle de la conservation du moment dans un univers avec homog\'en\'eit\'e
spatiale: une cons\'equence du PRINCIPE COSMOLOGIQUE de Milne-Einstein. On montre que c'est
d'accord avec la th\'eorie de relativit\'e g\'en\'erale d'Einstein (par l'application du
Principe d'Action). On appelle cet effet ``MASS BOOM''. Au syst\`{e}me de LABORATOIRE aucune
de ces changements du temps ne peuvent pas \^{e}tre d\'etect\'es, \`{a} moins que nous
fassions des comparaisons avec des observations cosmologiques (c.-à-d. en utilisant
diff\'erentes r\'ef\'erences loin du LABORATOIRE, par exemple mesurant le d\'ecalage rouge des
galaxies \'eloign\'ees en termes de normes de LABORATOIRE). On doit souligner que les
conditions physiques implicites par une masse variable avec le temps, comme la MASS BOOM
\'etablit, ainsi qu'une vitesse de lumi\`{e}re variable avec le temps, pr\'eservant la
constance du moment, sont compatibles avec les \'equations d'Einstein. On int\`{e}gre alors
ses \'equations cosmologiques et on trouve la solution pour le facteur d'\'echelle
cosmologique $a(t) = \text{constant} \cdot t^2$, ceci implique une expansion acc\'el\'er\'ee
apparente pour l'Univers, comme celui qui est vu depuis le LABORATOIRE. C'est
l'interpr\'etation donn\'ee aux observations r\'ecentes obtenues à partir de la Supernova Type
Ia. La détermination du facteur de d'\'echelle comme $a(t) = \text{constant} \cdot t^2$ est
bas\'ee sur une interpr\'etation de LABORATOIRE et est donc un effet apparent. D'autre part
nous notons que le produit $ct$  \'etant une constante d\'etermine l'univers vrai comme un
univers statique, de mesure constante, selon la premi\`{e}re proposition d'Einstein. Le
d\'ecalage rouge observ\'e au syst\`{e}me de LABORATOIRE est dû à un vrai r\'etr\'ecissement
du monde de quantum, dû à la diminution de la taille des particules de quantum d\'etermin\'ees
par la d\'ecroissante ``constant'' de Planck $\hbar$ (l'importance des quantum sont de l'ordre
$\frac{\hbar}{mc}$, o\`{u} le $mc$ reste constant).

\bigskip\noindent
Mots cl\'es: Cosmologie, principe d'action, constantes physiques fondamentaux, quanta de
gravit\'e, supernova, le principe du Mach, Relativité Générale.
\bigskip

\end{resum}


\section{INTRODUCTION}

The question of the possibility of existence of time-varying physical constants in Nature has
been very much addressed in the literature. In particular much work has been done on the
possibility that the gravitational constant $G$ may be varying with time. No significant
experimental evidence exists in support of such time variations in the fundamental physical
constants.  In this work we find that the gravitational masses, and the speed of light $c$,
consistent with Einstein's Theory of general Relativity, vary with time. Their product $mc$ is
constant, as required by the constancy of momentum (the most preserved principle in Nature).
We prove that a time-varying $c$ implies a time varying $G$ and viceversa. We apply the Action
Principle, Mach's principle, what we call the Total Interaction principle, and the emission
rate of gravity quanta, to find a universal effect that we call ``Mass Boom'', a linear
increase of mass with cosmological time and therefore a cosmological effect. Weinberg's
relation also supports this conclusion. The increase of mass with time (the Mass Boom) is
``compensated'' by a universal negative energy boom present in the background gravitational
field. This is a result, already present in the scientific work, implied in a ``free lunch''
idea of creation of the Universe. Polarization of the vacuum in the form of positive (Mass
Boom) and negative energy (the gravitational field) is the explanation for the gravity quanta
emission from particles$^3$ producing the universal gravitational attraction introduced by
Newton.

The Mass Boom$^1$ implies that the speed of light decreases with time, as required by the
constancy of momentum. We then integrate Einstein's cosmological equations under the
assumption of a cosmological scale factor $a(t)$ varying as $t^x$ (and for the flat case, $k =
0$). With the solution $\Omega_m=1/3$ and $\Omega_\Lambda=2/3$, i.e. a ratio
$\Omega_m/\Omega_\Lambda=L=2$ we find the value of $x$ to be 2, which implies an apparent
(from the LAB) accelerated expansion for the Universe. The ratio $L$ may have varied through
the history of the Universe implying deceleration or acceleration depending on its value at
each time. But the ``real'' Universe is of constant size, $ct = \text{constant}$, the red
shift being a result of the ``real'' contraction of the quantum world.

\section{THE ACTION INTEGRAL}

Einstein's field equations can be derived from an action integral following the Least Action
Principle. In standard General Relativity one has for the action integral$^2$:
\begin{equation}
 \begin{aligned}
     A&=I_G+I_M \\
     A&=-c^3/\left(16\pi G\right)\int R(g)^{1/2}\,d^4 x + I_M
 \end{aligned}
\end{equation}
where $I_M$ is the matter action and $I_G$ the gravitational term. Then one obtains the field
equations
\begin{equation}
G^{\mu\nu} = 8\pi\left(G/c^4\right)\cdot T^{\mu\nu}
\end{equation}
We assume a space-time metric and use the Robertson-Walker model that satisfies the Weyl
postulate and the cosmological principle, i.e.
\begin{equation}
 ds^2 = c(t)^2\, dt^2 - R(t)^2 \left\{ dr^2/\left(1-kr^2\right) + %
 r^2 \left( d\theta^2 + sin^2\theta \, d\phi^2 \right) \right\}
\end{equation}
Einstein's equations (2) follow from the Action (1) provided that  the variation of the
coefficient in the integral  in equation (1)  be zero. Then
\begin{equation}
c^3 / (16\pi G)  =  \text{constant}
\end{equation}
We see that the assumption of a time varying $G$ must include a time varying $c$ to preserve
the form of the field equations.

On the other hand, the action for a free material point is
\begin{equation}
  A=-m c \int ds
\end{equation}
To preserve standard mechanics we make the momentum $mc$ constant, independent from the
expansion of the Universe, then
 \begin{equation}
   mc = \text{constant}
 \end{equation}
This condition implies that any time variation in m requires a time variation in $c$. With the
constancies expressed in (4) and (6), General Relativity is preserved and of course the
Newtonian mechanics too. Within these limits time variations of some of the fundamental
constants, $G$, $c$ and masses, are allowed in the sense of preserving the laws of physics, as
we know them today. Einstein's field equations are therefore compatible with this time
variations provided that both (4) and (6) hold.

\section{MACH'S PRINCIPLE}

One way to have a mathematical expression for Mach's principle is to say that the
gravitational potential energy of any mass $m$, with respect to the mass $M_u$ of the rest of
the seeable Universe with size $ct$, is of the order of the relativistic energy of the mass
$mc^2$ i.e.
 \begin{equation}
   \frac{GM_u m}{ct} \approx mc^2
 \end{equation}
Hence, using the relation (4) we get in a certain system of units
 \begin{equation}
   G=c^3
 \end{equation}
and therefore substituting $G$ in (7) we finally have
 \begin{equation}
   M_u = t
 \end{equation}
This is what we call the Mass Boom$^1$. It implies that the mass of the Universe increases
linearly with time. If the number of particles in the seeable Universe is constant, then their
mass must be increasing linearly with time too, in accordance with the idea of the particles
emitting gravity quanta of negative mass$^3$. The resultant model for the Universe is then a
static one full of positive energy particles, increasing their mass with time (not the number
of particles that is a constant), together with the negative energy background of
gravitational potential that becomes more and more negative with time. This is a polarization
picture, a ``free lunch'' model for the Universe.  From (6) we see that the speed of light
must be decreasing with time. It is proportional to $1/t$ , an idea already advanced in many
works. Also an increase of the mass of the Universe with cosmological time has been postulated
in the past$^{4,5}$ by some authors like Dirac (1937) and Arnot (1941). However, their
postulated increase of mass for the Universe is due to particle creation. In our case particle
creation is not present. It is the mass $m$ of each particle that increases linearly with
time, giving the final result of a Mass Boom. Also, in our case this increase is responsible
for the gravitational force, through the emission of gravity quanta of negative mass$^3$ . And
the Mass Boom is compensated by the decreasing negative energy of the gravitational field due
to the addition of more and more negative gravity quanta. This result is in contrast with the
assumption made by Milne$^6$  in 1937 (and earlier papers). He derived the same relation (7)
and took the mass $M_u$ as well as  the speed of light $c$ as constants. Then he obtained $G$
proportional to $t$. Obviously this approach of Milne does not satisfy the condition in (4)
for $G$ to be proportional to $c^3$, a condition necessary to derive the Einstein's field
equations from the Action Principle. The difference between Milne's approach and ours is that
he postulated his own relativistic theory, different from the General Theory of Relativity of
Einstein. By contrast we keep Einstein's field equations as the valid ones for cosmology.

We present now what we call the Total Interaction Principle. It is a mathematical expression
that follows the requirement that all the gravitational interactions in the Universe must have
a mean free path, under a Newtonian point of view, of the order of the size of the Universe.
Then,
 \begin{equation}
   ct\approx\frac{1}{n\sigma_g}
 \end{equation}
where $n$ is the number density of particles in the Universe and $\sigma_g$ their
gravitational cross section as defined elsewhere$^3$ and given by
 \begin{equation}
  \sigma_g = 4\pi \frac{Gm}{c^2} \cdot ct
 \end{equation}
Substituting the above into (10) one has
 \begin{equation}
  ct\approx \dfrac{(ct)^3}{\dfrac{GM_u}{c^2}\,ct}
 \end{equation}
i.e.
 \begin{equation}
   \dfrac{GM_u}{c^2}\approx ct
 \end{equation}
which is the same as (7), Mach's Principle.

We therefore have 3 principles, Action Integral, Mach's and the Total Interaction, that
converge into the same conclusion. This is a strong theoretical drive to conclude that the
mass of the Universe, and therefore the mass of any fundamental particle, increases linearly
with time: the Mass Boom that we have proposed elsewhere$^1$.

Furthermore, by using the mass of the quantum of gravity $m_g$ defined elsewhere$^3$  as
 \begin{equation}
  \left|m_g\right|=\frac{\hbar}{c^2 t}
 \end{equation}
Let us call $m$ the mass of a typical quantum particle. The time $\tau$ for one quantum of
gravity of mass $m_g$ to be emitted by the particle of mass $m$ must be of the order of the
time taken for light to travel  a Compton size $\hbar/mc^2$. Calculating the mass rate of
change $dm/dt$ as given by the ratio $m_g/t$, one has:
 \begin{equation}
   \frac{dm}{dt}\approx\frac{\hbar}{c^2 t}\,\frac{mc^2}{\hbar} = \frac{m}{t}
 \end{equation}
and equivalently we obtain from this
\begin{equation*}
  \frac{dm}{m}\approx \frac{dt}{t}
\end{equation*}
This equation is immediately integrated to give the solution
 \begin{equation}
   m=\text{const}\cdot t
 \end{equation}
and therefore we obtain again the Mass Boom effect.

To increase the evidence in favor of the Mass Boom concept we refer now to Weinberg's
relation$^2$:
 \begin{equation}
   m=\left(\frac{\hbar^2}{Gct}\right)^{1/3}
 \end{equation}
One can see that this relation, that gives an order of magnitude for quantum masses $m$ in
terms of quantum mechanical, gravitational and relativity quantities ($\hbar$, $G$ and $c$),
as well as the cosmological time $t$, implies from (4), (and $\hbar$ = constant at the LAB),
$\left(m^3c^3\right)\cdot ct $= constant. Using now (6) we finally get the fundamental
relation $ct$ = constant. Since from (6) we have $mc$ = constant we also have the other
fundamental relation of this theory $m\propto t$, the Mass Boom again. We can also see that
Weinberg's relation (17) can be derived  by equating the square of the Compton wavelength of
the mass $m$ to the gravitational cross section (11).

\section{LABORATORY PHYSICS AND COSMOLOGY}

The key point when measurements of fundamental physical constants are made at the LAB, with no
reference to cosmological observation and/or cosmological parameters, is that we can not
escape of making relative comparisons, ratios of the same dimensional properties. Then there
is no way to detect any possible time variation of this ``constants'' because all measurements
are relative ones, lengths to lengths, mass-to-mass etc. If there is any time variation with
the same law for example for lengths, or for masses, the relative comparison makes the time
variation undetectable. This is a clear and well-known fact.

However, the well known theorem of the conservation of energy has to be reinterpreted now.
Energy, in our approach, varies with time as $mc^2 = mc\cdot c = \text{const}\cdot c \propto
1/t$ i.e. energy decreases with time. By Noether's theorem there would be no invariance under
time displacements, or equivalently the time coordinate would not be homogeneous (in a similar
way the homogeneity of space implies the conservation of momentum). Then, from a quantum
mechanical point of view, based upon Heisenberg's relation, energy x time should be constant
at the LAB (equal to Planck´s constant). We conclude that the time measured at the LAB is not
homogeneous, i.e., the tics from the clocks we use must be proportional to the cosmological
time $t$.

We can check this result. The atomic clocks have a tic proportional to $\hbar/mc^2$. Since
$\hbar$ is constant (LAB) and $mc$ is also constant (always) then the tic is inversely
proportional to $c$, i.e., directly proportional to time $t$. And the same happens if we use
gravitational clocks (periods of revolution $T$): by Kepler's third law  $T^2$ is proportional
to the cube of the size$r$ of the orbit divided by $GM$, that is $c^3 \cdot M =
\text{constant}\cdot c2$ (using again our relations in (4) and (6)). In fact the size of the
orbits is constant in our theory: $v^2\approx GM/r \propto c^2/r$. From Special Relativity
$v/c = \text{constant}$ and therefore $r = \text{constant}$. The conservation of angular
momentum, $mvr = \text{constant}$, gives the same result. And this is a very important one:
the sizes of the Newtonian orbits are constant. And so is the size of the Universe. Kepler's
third law gives then the result we have envisaged: periods are proportinal to the cosmological
time $t$. Planck's unit of time is also porportional to $t$ at the LAB, with a constant
$\hbar$. We have proved that masses increase linearly with time, and the speed of light
decreases accordingly. However, the experiments to measure the time variations of these
``constants'' will not succeed in seeing them. For example, the lunar ranging experiments
based upon lunar dynamics, that considers the relations:
 \begin{equation}
   \begin{aligned}
     v^2&=\frac{GM}{r} \\ mvr&= \text{const.}
   \end{aligned}
 \end{equation}
concludes that $G$ is practically constant. Now, the elimination of the speed $v$ implies the
following relation used in the analysis of lunar ranging data:
 \begin{equation}
   GMm^2r=\text{const}
 \end{equation}
This last relation gives, using $G = c^3$ and the masses $M$ and $m\propto t$, $c\propto 1/t$,
the final result of $r=$constant. This is what is observed: no time variation in $G$ because
this ``constant'' combines with the masses to give a real constant. The product $GMm^2$ is
obviously a constant in this theory. If we take the masses as constant, then $G$ is obviously
a constant. \emph{It is impossible to measure any time variation in $G$ in his way}\/.

\section{COSMOLOGICAL EQUATIONS}

The Einstein cosmological equations derived from his General Theory of Relativity are
 \begin{equation}
    \begin{aligned}
       \left(\frac{\dot a}{a}\right)^2+\frac{2\ddot a}{a}+8\pi\,G\frac{p}{c^2}+%
       \frac{kc^2}{a^2} &=\Lambda \,c^2\\ %
       \left(\frac{\dot a}{a}\right)^2-\frac{8\pi}{3}\,G\rho + \frac{kc^2}{a^2}%
       &=\frac{\Lambda\,c^2}{3}
    \end{aligned}
 \end{equation}
Using a flat Universe ($k = 0$), consistent with recent observations, and a zero pressure
Universe as given by an average zero thermal speed for galaxies (the atoms of the Universe)
one has
\begin{equation}
    \begin{aligned}
       &1+\frac{2\ddot a a}{\dot a^2} = 3\Omega_{\Lambda}\\ %
       &1=\Omega_m + \Omega_\Lambda
    \end{aligned}
 \end{equation}
where
\begin{equation}
    \begin{aligned}
       \Omega_{m}&=\frac{8\pi}{3}\,\frac{G\rho a^2}{\dot a^2}\\ %
       \Omega_\Lambda&=\frac{\Lambda c^2}{3}\,\frac{a^2}{\dot a^2}
    \end{aligned}
 \end{equation}
Let us assume a law for the cosmological scale factor $a$ of the form
 \begin{equation}
   a(t)=A\,t^x
 \end{equation}
where $A$ and $x$ are constants and $t$ the age of the Universe. Defining a parameter $L$ as
 \begin{equation}
  L=\frac{\Omega_m}{\Omega_\Lambda}
 \end{equation}
one derives from (21), (23)  and (24)
 \begin{equation}
  x=\frac{2}{3} \left(1+L\right)
 \end{equation}
Present numerical values for the parameters are very close to
 \begin{equation}
   \begin{aligned}
     \Omega_\Lambda&=\frac{2}{3}\\
     \Omega_m&=\frac{1}{3}
   \end{aligned}
 \end{equation}
i.e. $L = 2$, consistent with many observations.  Then one has from (28) $x = 2$ and therefore
$a(t) = At^2$.  The Universe is now accelerated, as observed from the Supernova Type Ia$^7,8$
results, with an acceleration parameter given by
 \begin{equation}
   q_0=-\frac{\ddot a \,a}{\dot a^2} = -\frac{1}{2}
 \end{equation}
The result is an apparent expansion of the Universe as $a(t) = At^2$. Since the ``real''
Universe is of constant size, $ct =$constant, the interpretation of the red shift is now that
the constant wavelengths of the photons coming from distant galaxies when compared with the
LAB reference we get the alternative result that it is the LAB wavelength that shrinks
($\hbar/mc$ is proportional to $\hbar$ and therefore the red shift observed, proportional to
$1/t^2$, gives $\hbar$ proportional to $1/t^2$). Hence de De Broglie wavelengths (sizes of
quantum particles) shrink like $\hbar$. And now we conclude that the ``natural or universal''
units of time and mass are the Planck's units of time and mass with the precise value h = c2 .
The ``natural or universal'' unit of length is of course the size of the Universe, $ct =$
constant $\approx 10^{28}$~{}cm.

\section{CONCLUSIONS}

We have presented here strong theoretical reasons, based upon fundamental principles, pointing
towards a ``Mass Boom'', a linear increase of all gravitational masses with cosmological time.
Correspondingly the speed of light decreases linearly with time. Hence the momentum is
conserved. The Einstein's cosmological equations, with the present values of the cosmological
parameters (flatness, mass, dark energy) predict an accelerated expansion of the Universe as
seen from the Lab system, in agreement with present day observations. One can not expect to
detect the increase of masses, or the decrease of the speed of light, from laboratory
experiments: the natural references used preclude the observation of these cosmological
effects. But the coherence of the approach presented here strongly suggests that the gravity
quanta emission mechanism, presented elsewhere$^3$, has a sound theoretical basis to be
considered as a link between General Relativity and Quantum Mechanics. Then, reinterpretation
of the red shift is necessary.

Since the product $ct$ is a constant, the immediate interpretation is that the model of the
Universe we are getting at is of constant size: no expansion is present at large scales.
However, at the Lab system we see a red shift from distant galaxies. We have to reinterpret
this red shift: there is no real expansion of the Universe, but there is a real contraction of
the quantum world as given by the decrease in Planck's constant. The constancy of the size of
the Universe obviously implies a constant wavelength for the Cosmic Microwave Background
Radiation. And the \emph{apparent}\/ expansion of the Universe follows a $t^2$ law, an
accelerated expansion (a constant one). Correspondingly, the quantum world is shrinking as
$1/t^2$2 due to the relation $\hbar=c^2=1/t^2$.

\section{REFERENCES}


\setlength{\itemsep}{-0.6 ex}   

\begin{itemize}
\item[[1]] A. Alfonso-Faus,  2001. \emph{Frontiers of Fundamental Physics 4}\/ page 309, Kluwer Academic. ``FFP 4''  Hyderabad, India, 2003 arXiv:physics/0302058 v1.
\item[[2]] S. Weinberg, 1972, \emph{Grav. and Cosm.}\/, (NY, John Wiley \& Sons).
\item[[3]] A. Alfonso-Faus, 1999, \emph{Physics Essays}\/, Vol. 12 N4 December, and ArXiv:gr-qc/0002065 v1.
\item[[4]] Dirac, P., \emph{Nature}\/, 139, 323 (1937).
\item[[5]] Arnot, F.L., \emph{Time and the Universe}\/, Aust. Med. Pub. Sidney, 1941.
\item[[6]] Milne, E.A., \emph{Proc. Roy, Soc. m A}\/, 158, 324 (1937).
\item[[7]] S. Perlmutter, et al., astro-ph/9812133, 1999, \emph{Ap J}\/, 517, 565.
\item[[8]] A. Riess et al. 1998, \emph{A J}\/, 116, 1009.
\end{itemize}


\end{document}